# Hazard Exposure Heterophily: A Latent Characteristic in Socio-Spatial Networks Influencing Community Resilience

Chia-Fu Liu[1*], Ali Mostafavi[2]

[1] Ph.D Student, Zachry Department of Civil and Environmental Engineering, Texas A&M University, 199 Spence St., College Station, TX 77843-3136; e-mail: joeyliu0324@tamu.edu
[2] Associate Professor, Zachry Department of Civil and Environmental Engineering, Texas A&M University, 199 Spence St., College Station, TX 77843-3136; e-mail: amostafavi@civil.tamu.edu

## Abstract

We present a latent characteristic in socio-spatial networks, hazard-exposure heterophily, to capture the extent to which populations with similar hazard exposure could assist each other through social ties. Heterophily is the tendency of unlike individuals to form social ties. Conversely, populations in spatial areas with significant hazard exposure similarity, homophily, would lack sufficient resourcefulness to aid each other to lessen the impact of hazards. In the context of the Houston metropolitan area, we use Meta's Social Connectedness data to construct a socio-spatial network in juxtaposition with flood exposure data from National Flood Hazard Layer to analyze flood hazard exposure of spatial areas. The results reveal the extent and spatial variation of hazard-exposure heterophily in the study area. Notably, the results show that lower-income areas have lower hazard-exposure heterophily possibly caused by income segregation and the tendency of affordable housing development to be located in flood zones. Less resourceful social ties due to high hazard-exposure homophily may inhibit low-income areas from better coping with hazard impacts and could contribute to their slower recovery. Overall, the results underscore the significance of characterizing hazard-exposure heterophily in socio-spatial networks to reveal community vulnerability and resilience to hazards.

**Keywords:** Community resilience; human networks; socio-spatial networks; Hazard-exposure heterophily

## Introduction

Examination of interactions within socio-spatial networks embedded in communities and social connections among disparate populations can illuminate subtle characteristics in community resilience assessments [1–5]. Socio-spatial networks represent the extent of interactions within spatial areas and social connections among populations of different areas. The intersection of socio-spatial network characteristics and spatial hazard exposure could reveal insights beyond the standard index-based approaches (such as social vulnerability index). This study reveals hazard exposure heterophily as a latent characteristic in socio-spatial networks of a community that could improve resilience and recovery. Heterophily refers to the tendency of people of groups to maintain a higher proportion of relations with members of groups other than their own [6]. Standard index-based approaches for spatial characterization of community vulnerability and hazard exposure do not capture the dynamics of populations and places in socio-spatial networks; and hence provide a limited view of the extent of a community's vulnerability and resilience. The identified hazard-exposure heterophily characteristic captures the extent to which residents in spatial areas exposed to natural hazards have social connections with residents outside hazard zones. The rationale is that social connections with non-hazard-prone areas are more resourceful during flood hazards. If residents of two spatial areas have strong social connectedness, but both experience flood impacts, they would probably not be resourceful to each other. In the context of this study, resourceful social ties are defined as social connections that can aid response and recovery from natural hazard impacts.

The existing literature has emphasized the importance of social cohesion and social capital in community resilience [7–11]. Researchers have proposed indices (such as social vulnerability index [12–15] and social capital index [16,17]) to specify the spatial patterns of social capital [18,19] and cohesion [10,11] across communities. However, the existing approaches suffer from two limitations: first, they determine the extent of social cohesion/capital based on attributes (e.g., socio-economic characteristics) of a spatial area (e.g., census tract or ZIP code tabulation area (ZCTA)) rather than on measured social connections [12,20–22]. This limitation is due mainly to challenges in measuring social connections among populations of different spatial areas. This limitation can be overcome using emerging data sources such as social media platforms. Second, existing approaches do not account for the extent of resourcefulness of social links during hazards by considering all social ties as homogenous. Residents of two spatial areas which are both flooded probably would not have resourcefulness to help each other. These two limitations have hindered the ability to examine the intersection of hazard exposure and socio-spatial networks simultaneously in an integrated manner.

In this study, we address this gap by examining hazard-exposure heterophily among spatial areas based on resourceful social ties (Figure 1). In communities, social connections are created based on homophily (the opposite of the characteristic of heterophily), similarity of socio-demographic attributes, such as income and race [23,24]. When natural hazards occur, these social connections are leveraged for responding and recovering from impacts of the hazard [25]. The rationale here is that the social connections with residents outside of a hazard zone are more resourceful during hazard events [9]. On the other hand, if residents of two spatial areas have strong social connection, but with both experiencing similar hazard impacts, the extent to which these social connections could be leveraged diminishes. This characteristic in socio-spatial networks can be explained based on hazard-exposure heterophily. In network science, heterophily in a network is defined as the extent of dissimilarity among the attributes of nodes that have links [6,26]. In socio-spatial networks, a greater hazard-exposure heterophily indicates a greater extent of resourceful social connections available to the population of a spatial area to aid them during their response and recovery from the impacts of a natural hazard. In the following sections, the datasets and methodology for assessing hazard-exposure heterophily are presented.

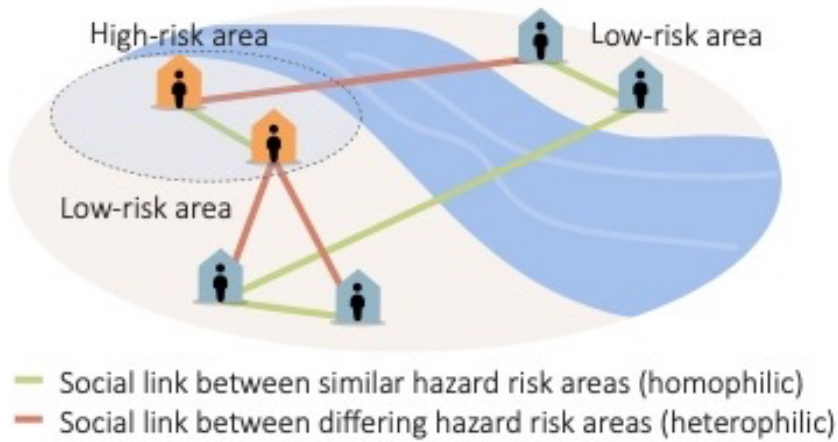

**Figure 1.** The illustration of *heterophily* and *homophily*.

## Data and Methodology

Harris County, which encompasses the Houston, Texas, metropolitan area was selected as the study area to demonstrate the approach for examining hazard-exposure heterophily. Harris County is one of the largest metropolitan areas in the United States and also one of the most flood-prone regions in the world. Hence, we selected this study area to examine hazard-exposure heterophily in the context of flood hazard exposure. We constructed the socio-spatial network model of community by using ZCTAs to represent spatial area nodes $i$ = 1, 2, …, n. The ZCTA geodatabase was extracted from the selected state-based geographic and cartographic information from the U.S. Census Bureau's Topologically Integrated Geographic Encoding and Referencing (TIGER) database. The geodatabases include feature class layers of information for the entire U.S.

To specify the level of hazard exposure of each spatial node (ZCTA), we focused on flooding as the primary hazard event in the study area. We collected 100-year and 500-year flood-hazard layer data from the Federal Emergency Management Agency (FEMA). FEMA provides the most recent nationwide extract of the National Flood Hazard Layer (NFHL) geospatial database through web-mapping services. This dataset includes flood map panel boundaries, flood hazard zone boundaries, and other information related to flood control zones and areas. The geospatial data was useful for calculating the floodplain area percentage (Figure 2) for each ZCTA and thus determine the level of flood hazard exposure. In this study, we included both 100-year and 500-year layers as the floodplain layer. We classified ZCTAs into two groups according to the median of floodplain-area percentage. ZCTAs with highest 50% floodplain area percentage are labeled as high-flood-exposure ZCTAs (H), while ZCTAs listed on lowest 50% floodplain area percentage would be regarded as low-flood-exposure ZCTAs (L). Accordingly, links between

ZCTAs with dissimilar hazard exposure attribute are designated as resourceful links and links between high-food-prone areas are designated as non-resourceful links.

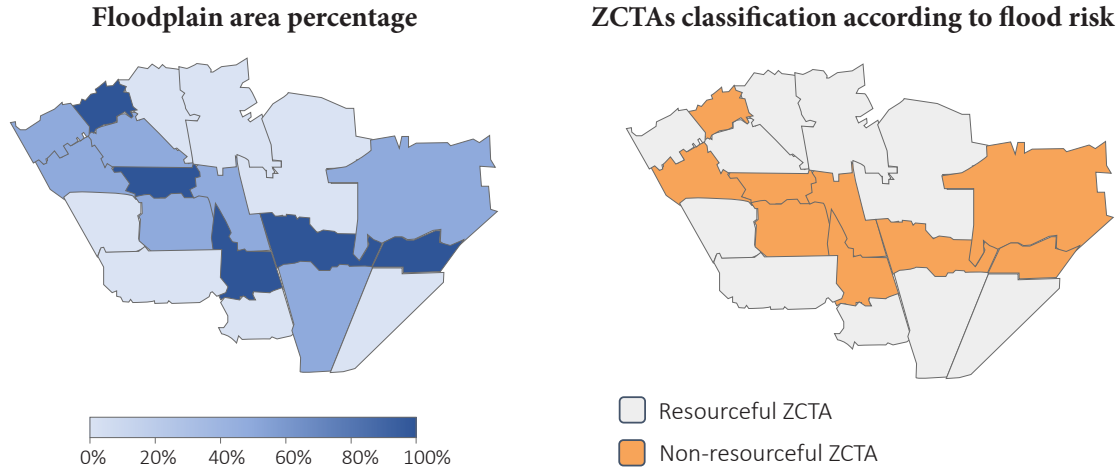

**Figure 2.** Left: The illustration of geographic distribution of floodplain area percentage. Right: The classification of the *High-hazard-exposure ZCTA* (*orange*) and *Low-hazard-exposure ZCTA* (*grey*).

To specify the *links* or *ties* that connect each spatial area node in socio-spatial network, we used the Social Connectedness Index (SCI) [27], which is based on friendship links on Meta, the company formerly known as Facebook, Inc. The SCI provided comprehensive measures of friendship networks at a national level. This dataset includes an aggregated and anonymized snapshot of all active Meta users; the relative frequency of Meta friendship links measures the intensity of connectedness between locations. Locations are assigned to users based on their information and activity on Meta, including the city of residence stated on their Meta profile, and device and connection information. In this study, SCI is presented at the ZCTA-level, which are pairs of five-digit ZCTA connections. Only ZCTAs with more than 500 residents and enough Meta users to produce meaningful estimates were included, leaving 26,271 of 33,139 ZCTAs covering 99.4% of the United States population. The measure of Social Connectedness $e_{i,j}$ between two ZCTAs *i* and *j* is

$$e_{i,j} = \frac{v_{i,j}}{User(i) \times User(j)} \quad (1)$$

where, $User(i)$ and $User(j)$ are the number of Meta users in locations *i* and location *j*, and $v_{i,j}$ is the number of Meta friendship connections between locations *i* and location *j*. For each value in the dataset, the measure was scaled to have a fixed maximum value by dividing the original measure by the maximum and multiplying by $10^9$ and the lowest possible value of 1. Each measure was rounded to the nearest integer.

*Methodology*

To capture hazard exposure heterophily, we defined the res_tie_rate $\rho_i$ ($i = 1, 2, …, n$) as a node attribute for each of ZCTA

$$\rho_i = \frac{\sum_{j \in L} e_{i,j}}{\sum_{j=1}^{n} e_{i,j}} \times 100\% \qquad (2)$$

The numerator is the summation of Social Connectedness that ZCTA *i* received from low-flood-exposure ZCTAs (*L*); meanwhile, the denominator is the total Social Connectedness that ZCTA *i* could receive from all its friendship ZCTAs (*L* and *H*). Notably, the fraction of Social Connectedness from low-flood-exposure ZCTAs (*L*) over the total Social Connectedness (*L* and *H*) aims to capture the degree of hazard exposure heterophily. Hazard-exposure heterophily exists in the friendship link which connects low-flood-exposure ZCTAs (*L*) with high-flood-exposure ZCTAs (*H*). On the other hand, a friendship link connecting two ZCTAs in the same group, which could be a *L–L* pair or an *H–H* pair, is said to display hazard-exposure homophily. Therefore, by calculating the resourceful tie rate for each node, we are able to quantify the degree of hazard exposure heterophily shown in the social-spatial network. Table 1 shows the summarized variable description in the network model.

**TABLE 1.** Variable description of the network model.

| Definition | Notation |
|---|---|
| Node of ZCTA | $i = 1, …, n$ |
| Social connectedness | $e_{i,j}$, $i, j = 1, …, n$ |
| Number of Meta friendship connections | $v_{i,j}$, $i, j \in N$ |
| Number of Meta users | $User(i)$, $i = 1, …, n$ |
| Low-flood-exposure ZCTAs | $L$ |
| High-flood-exposure ZCTAs | $H$ |
| Resourceful Tie Rate | $\rho_i$, $i = 1, …, n$ |

## Results

In this section we discuss the results of hazard exposure heterophily in the context of Harris County. First, we extracted the geographic data of 140 ZCTAs in Harris County from the TIGER geodatabase and created nodes representing the spatial network. Second, to measure the level of hazard exposure for the 140 ZCTAs, we combined both 100-year and 500-year flood hazard layers in Harris County and calculated the floodplain area percentage for each ZCTA (Figure 3). Then, we divided the 140 ZCTAs based on the median of floodplain area percentage (28.77%) into a high-flood-exposure group (*H*) and a low-flood-exposure group (*L*). Third, we constructed the socio-spatial network by adding the Social Connectedness links and weights to the network.

In total, Harris County contains 223 links among ZCTAs based on the Meta's Social Connectedness Index.

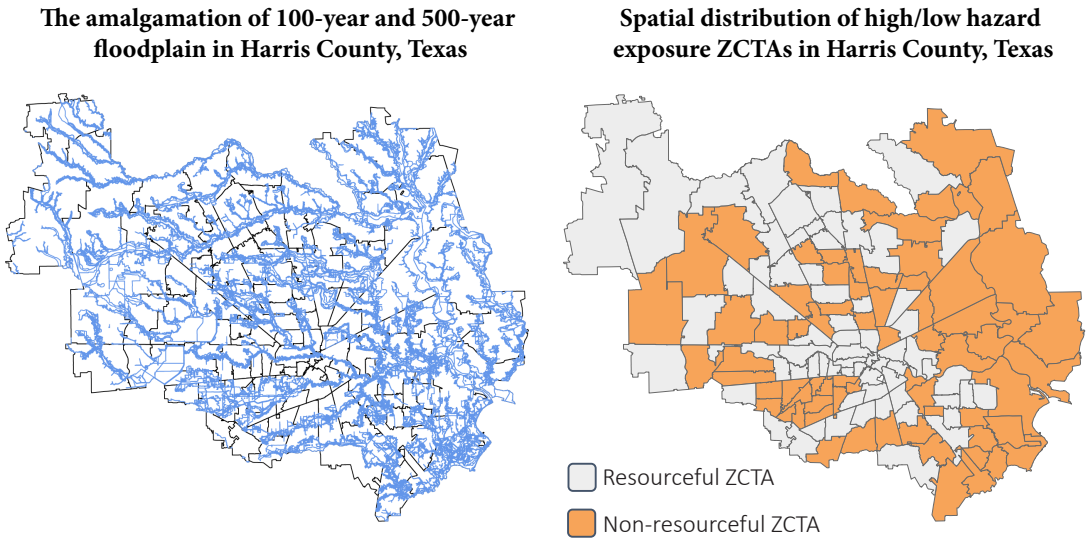

**Figure 3.** Left: The amalgamation of 100-year and 500-year floodplain in the Harris County, Texas. Right: The classification of high-hazard-exposure ZCTAs (*H*) and low-hazard-exposure ZCTAs (*L*) in Harris County, Texas.

In the next step, we calculated the res_tie_rate $\rho_i$ for 140 ZCTAs and compared it with the summation of Social Connectedness $\sum_{j=1}^{n} e_{i,j}$, which measures the total friendship link of ZCTA *i*, to calculate the extent of the resourcefulness of each link among node pairs. In Figure 4, the sum_of_SC shows a right-skewed unimodality, while the res_tie_rate appears in a more bimodal shape. The bimodal distribution shown in the res_tie_rate indicates the disparity of resourceful connections among 140 ZCTAs.

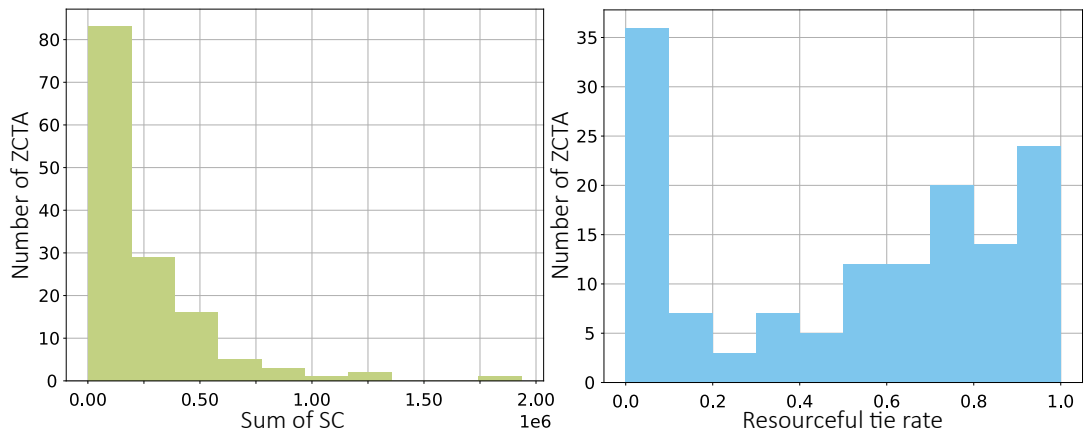

**Figure 4.** Left: The distribution of Sum of SC with skewness = 2.82. Right: The distribution of Resourceful tie rate with skewness = -0.26.

In the second step, 140 ZCTAs were classified into four groups based on two attributes: FP_rate and res_tie_rate (as shown in Figure 5 and Figure 6). We selected the median of FP_rate (28.77%) and the median of res_tie_rate (60.37%) to be the cutoff points for both classification ZCTAs with FP_rate less than 28.77%, fall into Group 1 (yellow) and Group 2 (green), include the low-flood-exposure ZCTAs (*L*). Group 3 (gray) and Group 4 (purple) include high-flood-exposure ZCTAs (*H*) (those with a FP_rate ≥ 28.77%). Among the low-flood-exposure ZCTAs, Group 3 has greater hazard exposure heterophily as indicated by the higher res_tie_rate. Hence, ZCTAs in Group 3 would have more resourceful links that could aid them during flood events, ostensibly enabling them to cope with the impacts and for a faster recovery. On the other hand, the lower res_tie_rate in Group 4 indicates hazard exposure homophily, which would reduce the ability of impacted residents in those areas to aid each other during flood events.

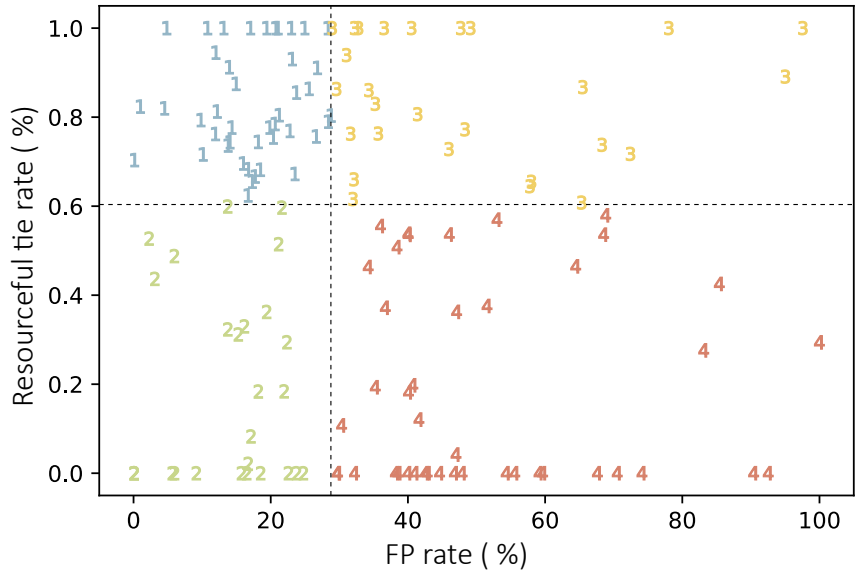

**Figure 5.** The classification of 140 ZCTAs in Harris County, Texas, according to FP rate and Resourceful tie rate.

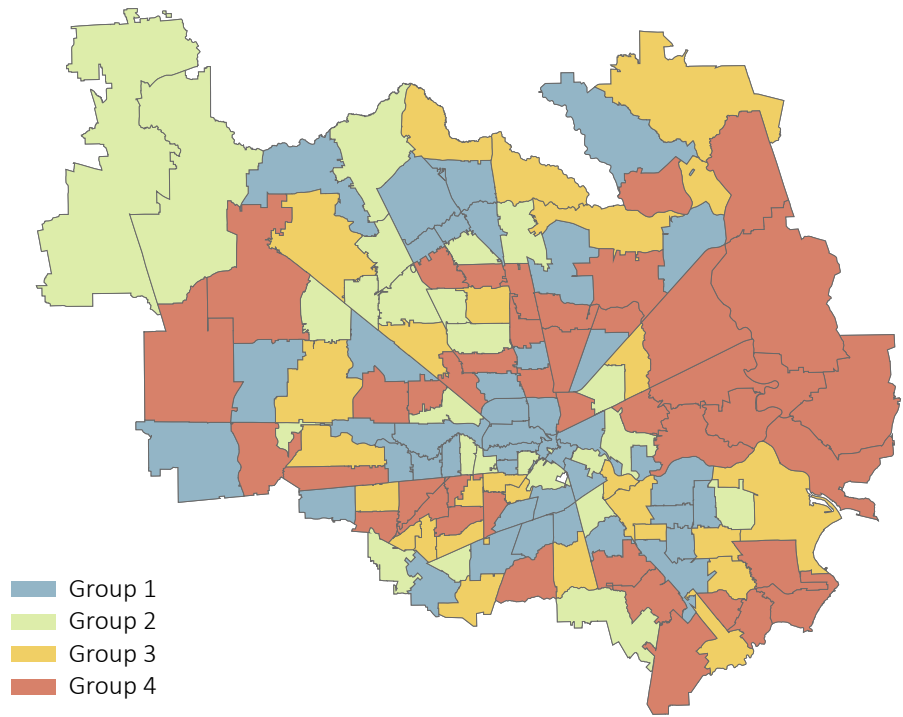

**Fig 6.** The spatial distribution of Group 1–Group 4.

Third, we evaluated the disparity in resourceful links in Group 3 and Group 4 with respect to their socio-economic information (shown in Table 2). We applied the statistical hypothesis test to the difference of median household income between Group 3 and Group 4. The result of the t-test shows that the means of median household income between Group 3 ($90.01K) and Group 4 ($67.61K) are significantly different (p-value = 0.026) with $\alpha = 0.05$ from each other. This implies that ZCTAs in Group 4 that show a lower degree of hazard exposure heterophily, and thus a higher degree of homophily, are mainly from lower income groups. This result is significant in two aspects: first, the results reveal one mechanism causing lower income areas recover more slowly after flood events. A number of studies [28,29] have reported the slow recovery of low-income areas after flood events in Harris County. The result presents one mechanism that negatively affects such slow recovery due to hazard exposure homophily. Second, this result highlights the effect of income segregation and community development patterns (locating affordable housing developments in flood zones[30,31]) on community resilience. The homophily among spatial areas with similar income levels when intersecting with the hazard exposure landscape would create a hazard-exposure homophily among these low-income areas which negatively affects their ability to cope and recovery from flood events.

TABLE 2. The mean of household income median for Group 1–Group 4.

| Group | Mean of household income median (in thousand) |
|---|---|
| 1 | $85.30 |
| 2 | $90.01 |
| 3 | $90.44 |
| 4 | $67.61 |

## Discussion

In this study, hazard exposure heterophily is introduced as a latent characteristic of socio-spatial networks that affects community resilience. Hazard-exposure heterophily captures the extent to which social connection among populations living in spatial areas with similar/dissimilar hazard exposure could be resourceful during hazard events, thus mitigating disaster impacts. The main idea in this study departs from the existing approaches in hazards/disaster research that consider the extent of social connections and hazard exposures in isolation and fail to capture the intersection of socio-spatial networks and hazard exposure in assessing community vulnerability and resilience. This study addresses this gap by revealing hazard exposure heterophily, which is a latent characteristic determining the extent of resourcefulness of social connections among spatial areas based on their respective hazard exposure.

This study leverages a unique social connectedness dataset based on Meta's Social Connectedness Index that provides a precise measure of social connections among residents of different spatial areas. This data is used in conjunction with the flood exposure data to construct socio-spatial network model of the community from which can be determined the extent of resourceful links associated to residents of each spatial area. Accordingly, a spatial area with low resourceful links can be identified and characterized as having a low hazard exposure heterophily (or high hazard exposure homophily). The findings reveal the spatial variation of hazard exposure heterophily. In addition, the findings reveal that low-income areas have lower hazard exposure heterophily (or greater hazard exposure homophily). The theoretical significance of this finding can be viewed from two aspects: first, this finding reveals one possible mechanism that hinders low-income areas from coping and recovering from flood impacts. The literature [28,29,32] provides strong evidence about greater impacts and slower recovery of low-income areas during flood events and has attributed the greater impacts and slower recovery to socio-economic characteristics [33–35]. The finding in this study however, reveals that hazard exposure homophily (or lack of heterophily) reduces the number of resourceful social links available to low-income areas, and thus could negatively affect their resilience and recovery. Second, these findings uncover that income segregation in communities (i.e., residents with similar income status have strong social ties), as well as affordable housing development in flood zones would lead to hazard exposure homophily in low-income areas, and subsequently adversely affect access to resourceful social ties. This result is consistent with the reported income segregation in cities and suggests that income segregation reinforces hazard exposure homophily and thus reduces the resourceful ties that low-income that can increase residents' need resilience and quicken recovery. Previous studies [30,31] have reported the negative effects of income segregation on communities; however, little theoretical explanation exists regarding the mechanisms by which income segregation affects community resilience and recovery. The finding from this study in the context of Harris County suggests that income segregation creates hazard exposure homophily, and thus, negatively affects access to resourceful social ties during hazard events. Furthermore, this study's outcomes open up new lines of inquiry for future studies to adopt the hazard-exposure heterophily characteristic and measures to explore the relationship between hazard-exposure heterophily and various aspects of community recovery and resilience. Accordingly, this study and findings contribute to and inform future research for better understanding of community resilience mechanisms at the intersection of socio-spatial networks and hazard exposure characteristics based on leveraging emerging datasets.

From a practical perspective, the specification of the spatial variation of hazard-exposure heterophily could complement the existing index-based approaches (such as the social vulnerability index and social capital index) to inform hazard mitigation, and response and recovery plans and actions. For example, the spatial areas with a lower number or degree of resourceful links can be prioritized by response and relief agencies and public officials for resource allocation and recovery assistance as they lack extensive resourceful social links.

**Data availability:**

The data that support the findings of this study are available from Meta Social Connectedness Data, but restrictions apply to the availability of these data. The data can be accessed upon request submitted on Meta Data for Good Program. Other data we use in this study are all publicly available.

**Code availability:**

The code that supports the findings of this study is available from the corresponding author upon request.

**Acknowledgment:**

The authors would like to acknowledge funding support from the National Science Foundation CAREER Award under grant number 1846069. The authors would also like to acknowledge the Meta Data for Good Program for providing the social connectedness data. Any opinions, findings, conclusions, or recommendations expressed in this research are those of the authors and do not necessarily reflect the view of the funding agencies.

**Author Contributions**

C.F.L. and A.M. conceived the idea. C.F.L collected the data and carried out the analyses. C.F.L. and A.M. wrote the manuscript.

**Competing interests**

The authors declare no competing interests.